\documentclass[a4paper,11pt]{article}
\pdfoutput=1

\usepackage{jcappub}

\title{\boldmath Higgs-portal assisted Higgs inflation with a sizeable tensor-to-scalar ratio}

\author[a,1]{Jinsu Kim\note{Corresponding author.}}
\author[a,b]{, Pyungwon Ko}
\author[c]{, and Wan-Il Park}

\affiliation[a]{Quantum Universe Center, Korea Institute for Advanced Study, 85 Hoegiro Dongdaemungu, Seoul 02455, Korea}
\affiliation[b]{School of Physics, Korea Institute for Advanced Study, 85 Hoegiro Dongdaemungu, Seoul 02455, Korea}
\affiliation[c]{Departament de F\'isica Te\`orica and IFIC, Universitat de Val\`encia-CSIC, E-46100, Burjassot, Spain}

\emailAdd{kimjinsu@kias.re.kr}
\emailAdd{pko@kias.re.kr}
\emailAdd{Wanil.Park@uv.es}

\abstract{
  We show that the Higgs portal interactions involving extra dark Higgs field can save generically the original Higgs inflation of the standard model (SM) from the problem of a deep non-SM vacuum in the SM Higgs potential.
  Specifically, we show that such interactions disconnect the top quark pole mass from inflationary observables and allow multi-dimensional parameter space to save the Higgs inflation, thanks to the additional parameters (the dark Higgs boson mass $m_{\phi}$, the mixing angle $\alpha$ between the SM Higgs $H$ and dark Higgs $\Phi$, and the mixed quartic coupling) affecting RG-running of the Higgs quartic coupling.
  The effect of Higgs portal interactions may lead to a larger tensor-to-scalar ratio, $0.08 \lesssim r \lesssim 0.1$, by adjusting relevant parameters in wide ranges of $\alpha$ and $m_{\phi}$, some region of which can be probed at future colliders.
  Performing a numerical analysis we find an allowed region of parameters, matching the latest Planck data.
}

\begin{document}

\begin{flushright}
KIAS-Q16017\,,
FTUV-16-10-11\,,
IFIC/16-69
\end{flushright}

	\maketitle
	\flushbottom

\section{Introduction}\label{sec:intro}
Even though inflation is a common sense for the very early history of the Universe, it is embarrassing to see that there are little compelling scenarios of inflation based on high energy physics theory.
Since a scalar field is needed in order to realize an observationally consistent inflation and due to the discovery of the standard model (SM) Higgs boson, the idea of using the SM Higgs field as an inflaton attained great attention.
Indeed it was shown that the SM Higgs field can play a role of inflaton if it has either a large nonminimal coupling to gravity \cite{Bezrukov:2007ep} or noncanonical kinetic term \cite{Germani:2010gm}.
Merits of the SM Higgs inflation include i) consistency with the latest Planck data \cite{Ade:2015xua,Ade:2015lrj}, ii) minimality of the model, and iii) the possibility of a connection between low energy (e.g., electroweak scale $m_{{\rm EW}}$) physics and high energy (e.g., inflation scale) physics \cite{DeSimone:2008ei,Hamada:2014iga,Bezrukov:2014bra}.

At tree level, with a large nonminimal coupling to gravity, the original SM Higgs inflation gives a small tensor-to-scalar ratio, $r\simeq 0.003$~\footnote{
	For the case of noncanonical kinetic term, readers may refer to ref.~\cite{Germani:2014hqa}.
	}.
While the result is in agreement with the latest Planck data ($r \lesssim 0.15$) \cite{Ade:2015lrj}, it is still an intriguing question whether or not a large $r$ ($\mathcal{O}(0.01 - 0.1)$) is possible, since it would be the first hint of the very high energy and early history of the Universe that is accessible in the near future.
A couple of groups, motivated by the BICEP2 announcement \cite{Ade:2014xna}, showed that the SM Higgs inflation can generate such a large $r$ at loop levels for a very precise choice of the SM Higgs boson and the top quark pole mass $M_{t}$.
However, the price to pay there was the vacuum instability (see e.g., refs.~\cite{Bezrukov:2014ipa, Buttazzo:2013uya, Degrassi:2012ry, Allison:2013uaa});
namely, the SM Higgs quartic coupling becomes negative due to radiative corrections.

In ref.~\cite{Haba:2014zda}, it was shown that Higgs portal interaction can save the Higgs inflation, where the benefit of Higgs portal interaction appeared only via loop contribution (see also ref.~\cite{Ballesteros:2015iua}).
There still exists a strong connection between $M_{t}$ and $r$, although it becomes somewhat milder than the case without the Higgs portal.
However, tree-level effect of Higgs portal interactions is generically much more significant if it exists, and it is important to investigate whether the Higgs inflation scenario assisted by Higgs portal interaction can have predictions for cosmological observables which are qualitatively different from those in the original Higgs inflation.
In particular it would be interesting to see if the extreme sensitivity on the top quark and the Higgs boson masses in the Higgs inflation scenario can be relaxed in the presence of Higgs portal interaction.

There is another way to drive inflation in the presence of Higgs portal interaction; namely, the extra scalar field, rather than the Higgs field, may play the role of inflaton as studied in refs.~\cite{Lerner:2009xg,Lerner:2011ge,Tenkanen:2016idg,Aravind:2015xst,Ballesteros:2016xej} (see also refs.~\cite{Kawai:2014gqa,Kawai:2015ryj} for supersymmetric extensions of the Higgs inflation).
We however will focus on the Higgs inflation here.

In this paper we point out that Higgs portal interaction, which is generic in hidden sector dark matter (DM) models \cite{hur_ko,Baek:2011aa,Baek:2012se,Baek:2012uj,Baek:2013dwa,Baek:2013qwa,Baek:2014goa,Ko:2014bka,Ko:2014nha}, revives the Higgs inflation by weakening the dependence on $M_{t}$.
Specifically we address the possibility of having a sizeable tensor mode $r \sim \mathcal{O}(0.01 - 0.1)$ without resorting to a specific value of $M_{t}$.
This is due to two additional parameters (the dark Higgs mass $m_{\phi}$ and the mixing angle $\alpha$ between dark Higgs and the SM Higgs bosons), in addition to the top quark mass $M_{t}$, that can control the running behavior of the Higgs quartic coupling $\lambda_{H}$.
We impose several constraints, such as the perturbative unitarity, DM direct detection, and DM relic density, in order to further constrain the mixing angle and the dark Higgs mass.

This paper is organized as follows.
We discuss the effect of Higgs portal interaction on the Higgs quartic coupling of SM in section~\ref{sec:Hportal}.
In section~\ref{sec:HpHi} the Higgs-portal assisted Higgs inflation is carefully studied.
We first argue generic features of the model and then demonstrate our arguments using a concrete model, namely the singlet fermion dark matter model, in section~\ref{sec:HpHiSFDM}.
Based on the latest Planck data we show an allowed parameter region.
Several other constraints on the model shall also be given before summarizing our results in section~\ref{sec:con}.

\section{Higgs portal interactions and scalar mixing}
\label{sec:Hportal}
The SM needs to be extended in order to accommodate at least nonzero neutrino masses, dark matter, baryo/leptogenesis, and inflation.
Especially, many beyond-the-SM (BSM) scenarios introduce a hidden sector, where DM resides, in addition to the visible SM sector.
Both sectors may communicate via gauge couplings and/or portal interactions as well as gravitational interaction.
Among them, Higgs portal interaction is particularly interesting because it allows not only thermal contact of hidden to visible sector but also channel of direct detection.
In addition, it can leave imprints on collider experiments.
Actually, the Higgs bilinear is the only SM operator which allows a renormalizable contact to hidden sector even for nonabelian dark gauge symmetry~\cite{hur_ko,Baek:2013dwa}.

In delivering our main point, among various Higgs portal interactions, it is enough to consider
\begin{align}\label{eqn:HiggsPortal}
  V \supset \lambda_{\Phi H} |\Phi|^{2} H^{\dagger}H
  \,,
\end{align}
where $H$ is the SM Higgs field and $\Phi$ can be either (real) singlet or multiplet charged under a hidden sector gauge group.
There may be large model-dependence in the hidden sector and in the potential of $\Phi$.
However, the effect of the Higgs portal interaction \eqref{eqn:HiggsPortal} on Higgs inflation is simplified to two cases:
$\langle \Phi \rangle = 0$
or
$\langle \Phi \rangle \neq 0$.
In the former case, the interaction gives an extra contribution to renormalization group (RG) equation of $\lambda_{H}$, and can push it above to avoid zero-crossing.
As a result, Higgs inflation becomes possible for a given $M_{t}$~\cite{Haba:2014zda}.
In the latter case, which has not been discussed in regard to Higgs inflation, there is a tree-level effect as well as the loop contribution \cite{Baek:2012uj}.
That is,
\begin{align}\label{eqn:lH-shifted}
  \lambda_{H} =
  \left[
  1 - \left(
  1 - \frac{m_{\phi}^{2}}{m_{h}^{2}}
  \right)
  \sin^{2}\alpha
  \right]
  \lambda_{H}^{{\rm SM}}
  \,,
\end{align}
where $\alpha$ is a mixing angle between $H$ and $\Phi$, $m_{\phi}$ ($m_{h}$) is the mass of the dark (SM) Higgs and $\lambda_{H}^{{\rm SM}} \equiv m_{h}^{2}/2v_{H}^{2}$ is the SM Higgs quartic coupling.
For $m_{\phi} > m_{h}$, the nonzero mixing angle $\alpha$ can easily remove vacuum instability along the SM Higgs direction even if loop contribution of $\lambda_{\Phi H}$ is not large enough.

Note that, for a given top quark pole mass, in case of loop contribution the renormalization scale $\mu_{{\rm min}}$ at which $d\lambda_{H}/d\ln\mu \approx 0$ does not vary much.
On the other hand, if $\lambda_{H}(\mu \sim m_{{\rm EW}})$ is shifted at tree level due to $\alpha$ and $m_{\phi}$ with a negligible loop-correction of $\lambda_{\Phi H}$, $\mu_{{\rm min}}$ can easily be pushed up to the Planck scale, and Higgs inflation can be realized without the problem of a deep non-SM vacuum.

\section{Higgs-portal assisted Higgs inflation}
\label{sec:HpHi}
Once the potential along the SM Higgs direction becomes monotonic, Higgs inflation becomes possible \cite{Bezrukov:2007ep}.
Let us first review the basics of Higgs inflation.
The relevant Lagrangian in the Jordan frame is given, in the unitary gauge, by
\begin{align}\label{eqn:L-jordan}
  \frac{\mathcal{L}}{\sqrt{-g}}
  =
  \frac{M_{{\rm P}}^{2}}{2}\left(
  1 + \xi_{h} \frac{h^{2}}{M_{{\rm P}}^{2}}
  \right)R
  +\mathcal{L}_{h}
  \,,
\end{align}
where $M_{{\rm P}}$ is the reduced Planck mass and $\mathcal{L}_{h}$ is the Lagrangian of the SM Higgs field only.
In the Einstein frame, obtained by the conformal transformation $g_{\mu\nu} \rightarrow \Omega^{2}g_{\mu\nu}$ with $\Omega^{2} \equiv 1+\xi_{h} h^{2}/M_{{\rm P}}^{2}$, the potential, in the limit of $h \gg M_{{\rm P}} / \sqrt{\xi_{h}}$, is given by
\begin{align}\label{eqn:V-E-can}
  U(\chi) =
  \frac{1}{\Omega^{4}}\frac{\lambda_{H}}{4}\left(
  h^{2} - v_{H}^{2}
  \right)^{2}
  \simeq
  \frac{\lambda_{H} M_{{\rm P}}^{4}}{4\xi_{h}^{2}}\left(
  1 - e^{-\frac{2\chi}{\sqrt{6} M_{{\rm P}}}}
  \right)^{2}
  \,,
\end{align}
where $v_{H} = 246\, {\rm GeV}$ is the vacuum expectation value of $h$ and $\chi$ is the canonically normalized field, which is related to $h$ by
\begin{align}
  \frac{d\chi}{dh} =
  \frac{\sqrt{1+(1+6\xi_{h})\xi_{h}h^{2}/M_{{\rm P}}^{2}}}{1+\xi_{h} h^{2}/M_{{\rm P}}^{2}}
  \,.
\end{align}
The Einstein-frame potential \eqref{eqn:V-E-can} is exponentially flat for $\chi \gg \sqrt{3/2}M_{{\rm P}}$, and can drive a slow-roll inflation with slow-roll parameters defined by
\begin{align}\label{eqn:epsiloneta}
  \epsilon =
  \frac{M_{{\rm P}}^{2}}{2}\left(
  \frac{U^{\prime}}{U}
  \right)^{2}
  \,,\quad
  \eta =
  M_{{\rm P}}^{2}\frac{U^{\prime\prime}}{U}
  \,,\quad
  \zeta^{2} =
  M_{{\rm P}}^{4}\frac{U^{\prime}U^{\prime\prime\prime}}{U^{2}}
  \,,
\end{align}
where `$\prime$' and `$\prime\prime$' represent respectively the first and second derivatives with respect to $\chi$.
Inflation ends when $\epsilon \sim 1$, and the number of $e$-foldings is given by
\begin{align}
  N_{e} =
  \frac{1}{M_{{\rm P}}^{2}}
  \int_{\chi_{f}}^{\chi_{i}} d\chi\,
  \frac{U}{U^{\prime}}
  =
  \frac{1}{M_{{\rm P}}^{2}}
  \int_{h_{f}}^{h_{i}} dh \,
  \frac{U}{dU/dh}\left(
  \frac{d\chi}{dh}
  \right)^{2}
  \,,
\end{align}
where subscript $i$ and $f$ stand for the specific time of interest during inflation and the end of inflation, respectively.
The power spectra of scalar and tensor perturbations are given by
\begin{align}
  P_{\mathcal{S}} =
                    \left(
                    \frac{H_{I}}{2\pi}
                    \right)^{2}\left(
                    \frac{\partial N_{e}}{\partial \chi}
                    \right)^{2}
                    \,,\qquad
  P_{T} =
          \frac{8}{M_{{\rm P}}^{2}}
          \left(
          \frac{H_{I}}{2\pi}
          \right)^{2}
          \,,         
\end{align}
where $H_{I}$ is the Hubble parameter during inflation.
The tensor-to-scalar ratio is therefore given by
\begin{align}\label{eqn:TSratio}
  r = \frac{P_{T}}{P_{\mathcal{S}}}
  \,.
\end{align}
Recent data from Planck satellite mission \cite{Ade:2015xua,Ade:2015lrj} showed
\begin{align}\label{eqn:PlanckPS}
	\ln(10^{10}P_{\mathcal{S}}) = 3.094 \pm 0.034\,,
	\quad
	\text{(68\%; TT,TE,EE$+$lowP)}
\end{align}
and
\begin{align}\label{eqn:PlanckCO}
	n_{s} = 0.9644^{+0.0095}_{-0.0096}
	\,,\quad
	r < 0.149
	\,,\quad
	\alpha_{s} = -0.0085^{+0.0147}_{-0.0154}
	\,,
	\quad
	\text{(95\%; TT,TE,EE$+$lowP)}
\end{align}
at the pivot scale $k_{*} = 0.05 \, {\rm Mpc}^{-1}$, where $n_{s}$ is the scalar spectral index, $\alpha_{s}\equiv dn_{s}/d\ln k$ is the running of the scalar spectral index.
Note that the upper bound on the tensor-to-scalar ratio \eqref{eqn:PlanckCO} is with the spectral running $\alpha_{s}$; in the absence of the running of the spectral index, the upper bound on $r$ becomes $r \lesssim 0.1$ \cite{Ade:2015lrj} (see also ref.~\cite{Kinney:2016qyl} for a recent analysis).

The original SM Higgs inflation \cite{Bezrukov:2007ep} predicted $r \simeq 0.003$ for $\xi_{h} \simeq 49000 \sqrt{\lambda_{H}^{{\rm SM}}}$ with $\lambda_{H}^{{\rm SM}} \sim 0.1$ which is well below the sensitivity accessible in the near future.
However, this prediction is a naive expectation since $\lambda_{H}$ during inflation would not be the same as its value at the low-energy scales (e.g., electroweak scale).
That is, one has to take RG running of $\lambda_{H}$ into account.
Noticing this aspect, several authors pointed out that RG-improved Higgs inflation may give a sizeable tensor-to-scalar ratio, $r \sim 0.1$, by adjusting the pole mass of top quark $M_{t}$ \cite{DeSimone:2008ei,Hamada:2014iga,Bezrukov:2014bra}.
It was also pointed out that loop contribution coming from Higgs portal interaction is useful to lift up the energy scale of Higgs inflation \cite{Haba:2014zda}.
As stated earlier, in the case of Higgs inflation in the pure SM, the exact value of $M_{t}$ is crucial.
For example, in order to achieve $r \sim \mathcal{O}(0.1)$, the top quark pole mass should be $M_{t} \simeq 171.5\,{\rm GeV}$ for $m_{h} \simeq 126\,{\rm GeV}$ \cite{Hamada:2014iga,Bezrukov:2014bra}.
However, recent analysis of data from various collider experiments showed that
\cite{Olive:2016xmw}
\begin{align}\label{eqn:toppolemass}
	M_{t} = 173.21 \pm 0.51 \pm 0.71\,{\rm GeV}\,,
\end{align}
indicating that Higgs inflation may be unlikely to occur within the SM.
However, if there exits Higgs portal interaction, a drastic change takes place, as we shall show in the following.

Generically Higgs portal interaction involves a new symmetry-breaking scalar field dubbed \textit{dark Higgs} here, and causes a tree-level mixing between the SM Higgs and the dark Higgs, as described in the previous section.
This mixing is of crucial importance in the physics of Higgs inflation, particularly with heavy $M_{t}$ which is potentially signaling vacuum instability.
When there is a mixing between the SM and dark Higgses, the tree-level SM Higgs quartic coupling is related to the physical mass of the SM-like Higgs by eq.~\eqref{eqn:lH-shifted}.
In addition, in the presence of the Higgs portal interaction \eqref{eqn:HiggsPortal}, the RG equation for the coupling $\lambda_{H}$ has an additional contribution coming from $\lambda_{\Phi H}$ compared to the case of the SM (see appendix \ref{apdx:RGEs} for details).
If $\lambda_{\Phi H}$ has a negligible effect on RG-running of $\lambda_{H}$, but only becomes the source of the Higgs mixing, then the running of $\lambda_{H}$ is essentially the same as the SM, except that the low energy boundary value of $\lambda_{H}$ is determined not only by $m_{h}$ and $v_{H}$ but also by $\alpha$ and $m_{\phi}$.
A proper choice of ($\alpha,m_{\phi}$) can make $\lambda_{H}$ to be positive up to the Planck scale (see e.g., ref.~\cite{Baek:2012uj}).
In this case, for $M_{t} \sim 173.2\,{\rm GeV}$, $\mu_{{\rm min}}$ appears to be larger than the Planck scale, and Higgs inflation requires a very large $\xi_{h}$ (e.g., $\mathcal{O}(10^{3-4})$) resulting in $r \ll \mathcal{O}(0.1)$.
On the contrary, if $\lambda_{\Phi H}$ is not negligible relative to the other SM contributions in the RG running of $\lambda_{H}$, $\mu_{{\rm min}}$ can then be adjusted.
In this case, we can control $\alpha$ and $m_{\phi}$ as well as $\lambda_{\Phi H}$ in order to have $\lambda_{H}(\mu_{{\rm min}})$ arbitrarily small with $\mu_{{\rm min}}$ well below the Planck scale, and $r \sim \mathcal{O}(0.1)$ can be realized.

Let us sketch the impact of RG running of the Higgs quartic coupling $\lambda_{H}$ on cosmological observables, focusing especially on values of the tensor-to-scalar ratio \eqref{eqn:TSratio}.
In order to take quantum corrections into account we quantize the theory in the Jordan frame~\footnote{
See ref.~\cite{George:2015nza} for the analysis of the SM Higgs inflation in the Einstein frame.
}, where the RG-improved effective action (see e.g., ref. \cite{Sher:1988mj} for a review) is given by
\begin{align}\label{eqn:RGimprovEA}
  \Gamma = \int d^{4}x\,\sqrt{-g}\left[
  \frac{M_{{\rm P}}^{2}}{2}
  \Omega^{2}R
  -\frac{1}{2}g^{\mu\nu}G^{2}
  \partial_{\mu}h\partial_{\nu}h
  -V+\cdots
  \right]\,,
\end{align}
where ``$\cdots$'' include the remaining subdominant terms, and
\begin{align}\label{eqn:RGimprovEV}
	V(t)
	&=
	\frac{\lambda(t)}{4}
	G^{4}(t)h^{4}(t)
	\,,\quad
	\Omega^{2}(t)
	=
	1+\xi_{h}(t)G^{2}(t)\frac{h^{2}(t)}{M_{{\rm P}}^{2}}
	\,,\quad
	G(t)
	=
	e^{
		-\int dt\, 
		\frac{\gamma}{1+\gamma}
	}
	\,,
\end{align}
with $\gamma(t)$ being the Higgs field anomalous dimension. Here $t=\ln(\mu/M_{t})$ with $\mu$ being the renormalization scale.
In the following the renormalization scale is chosen to be $\mu = (\lambda_{t}/\sqrt{2})h$, where $\lambda_{t}$ is the top Yukawa coupling, so that the radiative correction is minimized~\cite{Bezrukov:2014bra}.

The observables are easily computed in the Einstein frame, where the potential \eqref{eqn:V-E-can}, ignoring the irrelevant vacuum expectation value $v_{H}$, takes
\begin{align}
  U \approx \frac{\lambda_{H}(t)G^{4}(t)h^{4}(t)}{4(1+\xi_{h}(t)G^{2}(t)h^{2}(t)/M_{{\rm P}}^{2})^{2}}\,.
\end{align}
The RG runnings of the parameters depend on a specific model under consideration.
Depending on models, the effect of $G(t)$ and the scale dependence of $\xi_{h}(t)$ may be ignorable (see e.g., figure~\ref{fig:NM-RG}).
In such a case as an illustration it is straightforward to see the impact of RG running of the Higgs quartic coupling $\lambda_{H}(t)$ on cosmological observables.
In the remaining of this section, therefore, let us focus only on the RG running of the Higgs quartic coupling $\lambda_{H}$ and discuss the general behavior of observables when quantum corrections are taken into account, before considering full RG runnings with a concrete model in the next section.

The slow-roll parameters \eqref{eqn:epsiloneta} can be expressed, in the large field limit, as follows:
\begin{align}\label{eqn:epsilonapprox}
  \epsilon \simeq
  \frac{4}{3\xi_{h}^{2}}
  \left(
  	\frac{M_{{\rm P}}}{h}
  \right)^{4}
  \left[
  	1+x \frac{\xi_{h}h^{2}}{4M_{{\rm P}}^{2}}
  \right]^{2}
  \,,\qquad
  \eta \simeq
  -\frac{4}{3\xi_{h}}\left(
  	\frac{M_{{\rm P}}}{h}
  \right)^{2}\left[
  	1 - \frac{3}{4}x - \frac{1}{4}xy
  \right]\,,
\end{align}
where
\begin{align}\label{eqn:xydefn}
	x \equiv \frac{1}{\lambda_{H}}\frac{d\lambda_{H}}{dt}
	\qquad\text{and}\qquad
	y \equiv \frac{1}{d\lambda_{H}/dt}\frac{d^{2}\lambda_{H}}{dt^{2}}
	\,.
\end{align}
The spectral index $n_{s}$ and the tensor-to-scalar ratio $r$ are then given by \cite{Stewart:1993bc,Liddle:1994dx,Leach:2002ar}
\begin{align}
	n_s
	&\approx
	1-6\epsilon+2\eta
	-\frac{2}{3}(5+36c)\epsilon^{2}
	+2(-1+8c)\epsilon \eta
	+\frac{2}{3}\eta^{2}
	+\left(
		\frac{2}{3}-2c
	\right)\zeta^{2}
	\,,\\
	r
	&\approx
	16\epsilon \left[
		1 + \left(
			-\frac{4}{3}+4c
		\right)\epsilon 
		+\left(
			\frac{2}{3}-2c
		\right)\eta
	\right]
	\,,
\end{align}
where $c = \gamma+\ln 2-2$ with $\gamma \approx 0.5772$ being the Euler-Mascheroni constant.
Note that we have included terms up to the second order of the slow-roll parameters. This next-order slow-roll correction is important to distinguish between the Higgs inflation and its variants as pointed out in ref.~\cite{Gorbunov:2012ns}.
In the classical limit, i.e., $x\rightarrow 0$ and $y\rightarrow 0$, we recover the usual relation, $\epsilon \simeq 3\eta^{2}/4$, as well as $r \approx 0.0032$ and $n_{s} \approx 0.966$.

On the other hand, it is certainly possible for the term with $x\xi_{h}h^{2}/M_{{\rm P}}^{2}$ in eq.~\eqref{eqn:epsilonapprox} to play some roles at cosmological scales of interest.
In this case, the tensor-to-scalar ratio can be large due to the RG effects.
Note that $d^{2}\lambda_{H}/dt^{2}$ depends on the beta-function of $\lambda_{\Phi H}$ as well as $\lambda_{\Phi H}$ itself.
Therefore the RG-running of $\lambda_{\Phi H}$ becomes also crucial, and it can be controlled by model-dependent parameter(s).

\begin{figure}[t]
\centering
\includegraphics[width=0.65\textwidth]{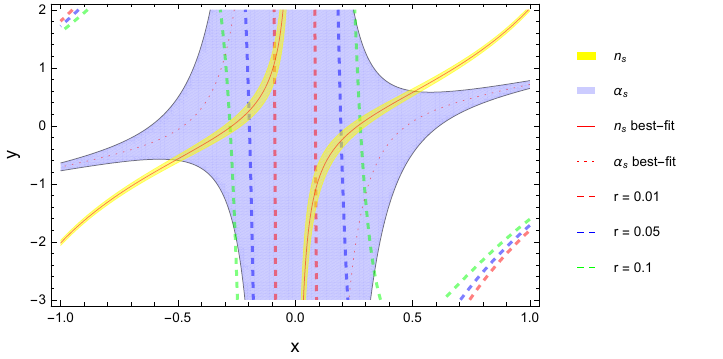}
\caption{\label{fig:paramspace}
Parameter space for cosmological observables. $\xi_{h}=400$ is chosen and $d^{3}\lambda_{H}/dt^{3}=0$ is assumed. Blue (yellow) colored region is the Planck constraint \eqref{eqn:PlanckCO} of the running of the scalar spectral index $\alpha_{s}$ (scalar spectral index $n_{s}$) with 95\% C.L. Solid (dotted) red curve is the best-fit value for $n_{s}$ ($\alpha_{s}$). Red, blue, and green dashed lines are respectively $r=0.01$, $0.05$, and $0.1$.
}
\end{figure}

In figure~\ref{fig:paramspace}, we show cosmological observables based on the latest Planck results \eqref{eqn:PlanckCO}.
A sizeable tensor-to-scalar ratio, $r\sim\mathcal{O}(0.1)$, may be obtained when
\begin{align}\label{eqn:condxy}
	|x| \approx 0.3
	\qquad\text{and}\qquad
	|y| \approx 0.1
	\,.
\end{align}
For a concrete model we shall consider below, where two-loop corrections to $\lambda_{H}$ and other couplings are taken into account, these conditions \eqref{eqn:condxy} are indeed well satisfied (see table~\ref{tab:InfObs}).

\section{Higgs-portal assisted Higgs inflation in the singlet fermion dark matter model}
\label{sec:HpHiSFDM}
Let us now demonstrate our argument with a concrete model.
In order to use the definite full RG-equations involving $\lambda_{H}$ and $\lambda_{\Phi H}$, we choose to work with the singlet fermion dark matter (SFDM) model \cite{Baek:2011aa,Baek:2012uj}.
In the SFDM model, the hidden sector consists of a singlet scalar field $S$ and a fermionic DM-candidate field $\psi$.
The hidden sector and the SM sector communicates through the following Higgs portal interaction:
\footnote{
	Although it is clear from the context, let us inform that $\Phi$ and $\lambda_{\Phi H}$ introduced in section~\ref{sec:Hportal} correspond to $S$ and $\lambda_{S H}/2$ in SFDM, respectively. In order to make notations clear we shall use $m_{s}$ instead of $m_{\phi}$ for the dark Higgs mass.
	}
\begin{align}
	V_{{\rm portal}}
	=
	\mu_{SH}SH^{\dagger}H
	+\frac{1}{2}\lambda_{SH}S^{2}H^{\dagger}H
	\,.
\end{align}
As stated in section~\ref{sec:HpHi}, we work in the Jordan frame in order to obtain the RG equations and then go to the Einstein frame where it is easy to study cosmological observables.
The full RG equations obtained up to two-loop order are summarized in appendix \ref{apdx:RGEs}.

Our numerical analysis is done as follows.
The RG equations \eqref{eqn:fullRGEs} are solved with the initial conditions at $\mu = M_{t}$ up to the Planck scale.
In order to determine the values of the $\overline{{\rm MS}}$ running parameters at $M_{t}$ scale, we used the C++ program library \texttt{mr} \cite{Kniehl:2016enc,Kniehl:2015nwa} which takes into account full two-loop threshold corrections and the full three-loop RG equations, together with the latest PDG values~\cite{Olive:2016xmw} at $M_{Z}$,
\begin{gather}
	M_{W} = 80.385\,{\rm GeV}\,,\quad
	M_{Z} = 91.1876\,{\rm GeV}\,,\quad
	M_{H} = 125.09\,{\rm GeV}\,,\quad
	M_{t} = 173.21\,{\rm GeV}\,,\nonumber\\
	G_{F} = 1.1663787\times 10^{-5}
	\,,\quad
	\alpha = 1/127.950
	\,,\quad
	\alpha_{s} = 0.1182
	\,,
\end{gather}
where $G_{F}$ is the Fermi constant, $\alpha_{s}(\mu) = g_{s}^{2}(\mu)/(4\pi)$ is the $\overline{\text{MS}}$ strong coupling structure constant and $M_{i}$ ($i=W,Z,H,t$) are the pole masses.
The new physics model parameters ($\lambda_{S}$, $\lambda_{SH}$, $\lambda_{\psi}$) as well as the nonminimal coupling ($\xi_{h}$) are chosen at $M_{t}$.
Note that $\xi_{h}$ is not a free parameter; we choose $\xi_{h}$ at the scale $M_{t}$ in such a way that the Planck normalization \eqref{eqn:PlanckPS} is satisfied.
Finally, we compute the slow-roll parameters and cosmological observables numerically, taking into account all the RG runnings, including $\xi_{h}(t)$, $\xi_{s}(t)$ and $G(t)$.

\begin{figure}[t]
\centering
\includegraphics[width=0.475\textwidth]{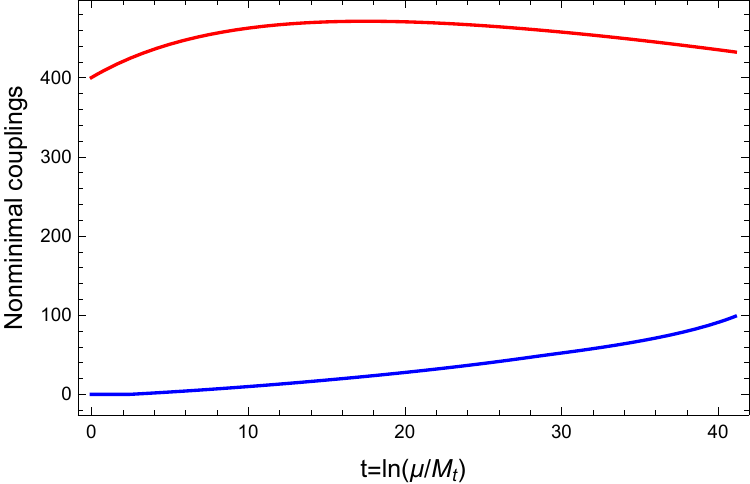}
\;\;\;
\includegraphics[width=0.475\textwidth]{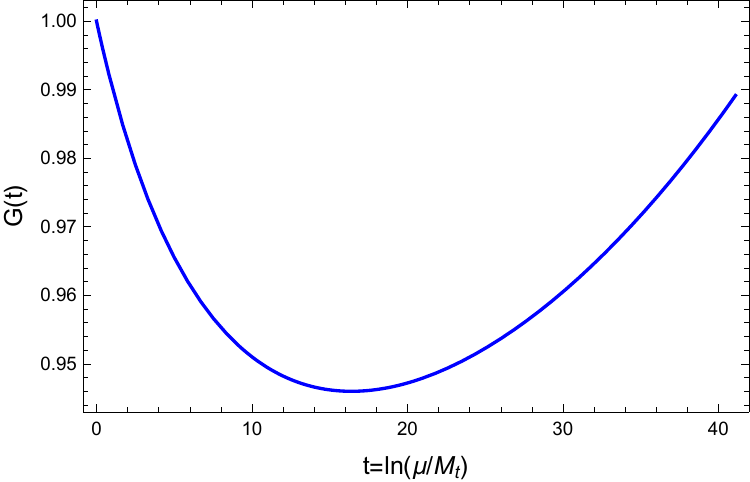}
\caption{\label{fig:NM-RG}
RG runnings of nonminimal couplings $\xi_{h}$ and $\xi_{s}$ and $G(t)$ \eqref{eqn:RGimprovEV} in SFDM. The initial conditions are chosen at $M_{t}$ scale as $\xi_{h} = 400$, $\xi_{s}=0$, $\alpha = 0.03$, $m_{s} = 500 \,{\rm GeV}$, $\lambda_{SH} = 0.1$, $\lambda_S = 0.2$, and $\lambda_\psi = 0.3$, where $\lambda_{S}$ is the quartic coupling of the extra singlet scalar $S$ and $\lambda_{\psi}$ is the coupling between the extra DM fermion $\psi$ and $S$, i.e., $\lambda_{\psi}S\overline{\psi}\psi$.
}
\end{figure}

It is important to note that a nonminimal coupling of the singlet scalar field $S$ to gravity, i.e.,
$	\mathcal{L} \supset 
	\tfrac{1}{2}\xi_{s}S^{2}R
	\,,
$
is generated via RG runnings.
Since we assume inflation to take place along the SM Higgs field, we set $\xi_{s}$ to be zero at $M_{t}$ scale and $S=0$ during inflation.
Figure~\ref{fig:NM-RG} shows the running of nonminimal couplings $\xi_{h}$ and $\xi_{s}$.

\begin{figure}[t]
\centering
\includegraphics[width=0.475\textwidth]{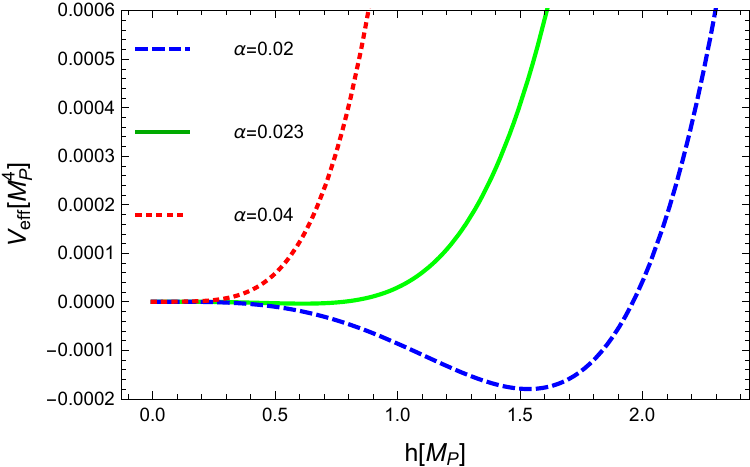}\;\;\;
\includegraphics[width=0.475\textwidth]{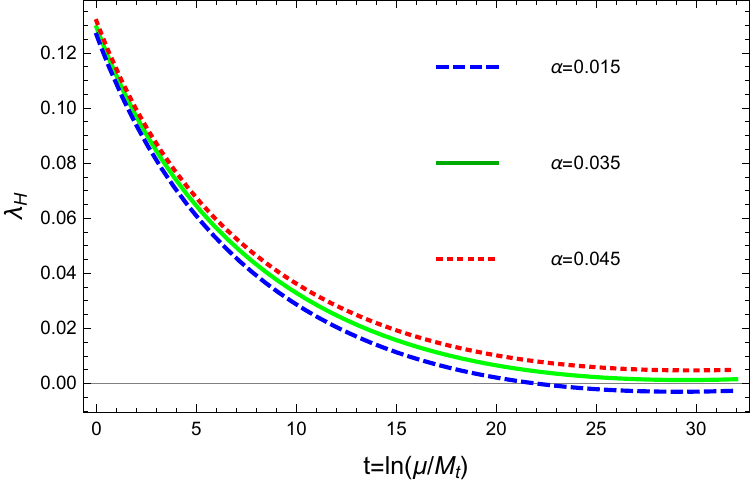}
\caption{\label{fig:Vhiggs}
Jordan-frame Higgs potential $V_{{\rm eff}}$ (left panel) and the running of $\lambda_{H}$ (right panel) in SFDM for $\xi_{h} = 440$, $\xi_{s}=0$, $m_{s} = 600 \,{\rm GeV}$, $\lambda_{SH} = 0.1$, $\lambda_S = 0.2$, and $\lambda_\psi = 0.3$ chosen at $M_{t}$ scale.
}
\end{figure}

In figure~\ref{fig:Vhiggs}, we show how the Jordan-frame RG-improved effective Higgs potential $V$ \eqref{eqn:RGimprovEV} and RG running of the Higgs quartic coupling $\lambda_{H}$ depend on the mixing angle $\alpha$ for fixed values of $m_{s}$ and $\lambda_{SH}$.
It is clear that, for a given $m_{s}$, the instability is determined by delicate interplay between $\alpha$ and $\lambda_{S H}$.
Note that one can achieve nearly the same behavior of the Higgs potential by adjusting $m_{s}$ instead of $\alpha$.
Therefore one may easily avoid the vacuum instability due to the presence of additional model parameters, while generating a large value of tensor-to-scalar ratio $r\sim \mathcal{O}(0.01-0.1)$ at the same time.
In other words, for a given value of top quark pole mass $M_{t}\sim 173.2\,{\rm GeV}$, the vacuum instability may be avoided once the mixing angle takes nonzero value, e.g., $\alpha \gtrsim 0.023$ in the case of figure~\ref{fig:Vhiggs}.

The $e$-foldings associated with a cosmological scale $\lambda = 2\pi/k$ is given by \cite{Liddle:1993fq}
\begin{align}
  N
  =
  62
  -\ln\left(
  \frac{k}{a_{0}H_{0}}
  \right)
  -\ln\left(
  \frac{10^{16}\,{\rm GeV}}{U_{I}^{1/4}}
  \right)
  +\ln\left(
  \frac{U_{I}^{1/4}}{U_{{\rm end}}^{1/4}}
  \right)
  -\frac{1}{3}\ln\left(
  \frac{U_{{\rm end}}^{1/4}}{\rho_{{\rm R}}^{1/4}}
  \right)
  \,,
\end{align}
where $\rho_{{\rm R}} = (\pi^{2}/30)g_{*}T_{{\rm R}}^{4}$, $g_{*}\approx 100$, $T_{{\rm R}}$ is the reheating temperature, $H_{0}\approx 0.67/(3000\,{\rm Mpc}$ is the Hubble parameter today \cite{Ade:2015xua}, and $U_{I}$ ($U_{{\rm end}}$) is the Einstein-frame potential at the horizon crossing (end of inflation).
For the SM Higgs inflation the reheating temperature is $T_{{\rm R}} \approx 1.8 \times 10^{14}\,{\rm GeV}$ \cite{Bezrukov:2014ipa}.
It is also pointed out in ref.~\cite{Bezrukov:2014ipa} that the upper bound on the reheating temperature is given by $T_{{\rm R}} \lesssim 5 \times 10^{15}\,{\rm GeV}$ which corresponds to the ``critical'' Higgs inflation case.
We choose $T_{{\rm R}} \approx 1.0 \times 10^{15}\,{\rm GeV}$ for our numerical analysis presented in table~\ref{tab:InfObs} since the values used are near the critical point.
Let us comment that the exact value of $T_{{\rm R}}$ barely alters our results in the sense that all the cosmological observables are in consistent with the latest Planck results.
\begin{figure}[t]
\centering
\includegraphics[width=0.65\textwidth]{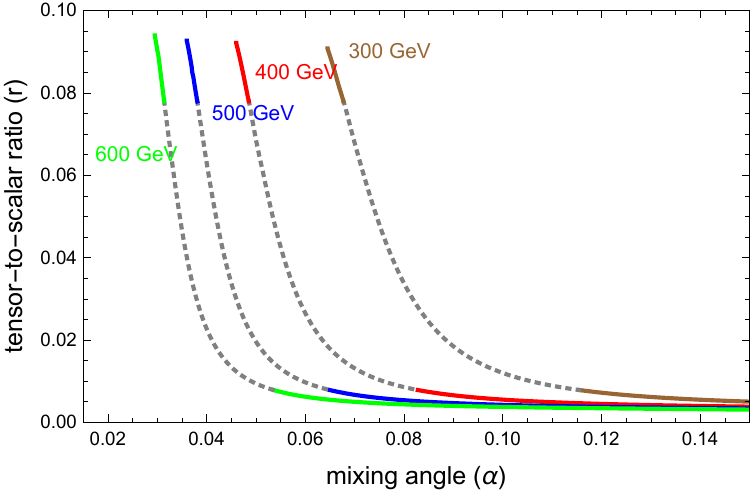}
\caption{\label{fig:paramscan}
Tensor-to-scalar ratio as a function of the mixing angle $\alpha$ for $m_{s}=300\,{\rm GeV}$, $400\,{\rm GeV}$, $500\,{\rm GeV}$ and $600\,{\rm GeV}$ at the pivot scale $k_{*}=0.05\,{\rm Mpc^{-1}}$. Here $\xi_{s}=0$, $\lambda_{SH} = 0.1$, $\lambda_S = 0.2$, and $\lambda_\psi = 0.3$ at $M_{t}$ scale are used. The nonminimal coupling of the SM Higgs to gravity, $\xi_{h}$, is chosen in such a way that the Planck normalization \eqref{eqn:PlanckPS} is satisfied. The grey-dotted lines indicate the parameter region where the spectral index $n_{s}$ becomes larger than $2\sigma$ Planck bound, $n_{s} \gtrsim 0.98$. Similar behaviors are found for different sets of model parameters.
}
\end{figure}
\begin{table}[t]
	\begin{center}
		\begin{tabular}{|c|c|c|c|c|c||c|c|c|c|c|}
			\hline
			$\alpha$ & $m_{s}$ & $\lambda_{SH}$ & $\lambda_{S}$ & $\lambda_{\psi}$ & $\xi_{h}$ & $N_e$ & $10^9 P_S$ & $n_s$ & $r$ & $\alpha_{s}$\\
			\hline \hline
			$0.036$ & $500$ & $0.1$ & $0.2$ & $0.3$ & $433$ & $57.3$ & $2.2$ & $0.9758$ & $0.0926$ & $-0.0003$ \\
			\hline
			$0.03885$ & $500$ & $0.1$ & $0.1$ & $0.1$ & $396$ & $57.3$ & $2.2$ & $0.9775$ & $0.0878$ & $-0.0003$ \\
			\hline
		\end{tabular}
	\end{center}
	\caption{\label{tab:InfObs} 
		Cosmological observables in SFDM. Two parameter sets which result in a sizeable value of the tensor-to-scalar ratio $r$ are presented. Here the pivot scale $k_{*}=0.05\,{\rm Mpc^{-1}}$ is chosen.
		For the upper (lower) case, we obtained $x \approx 0.25\;(0.26)$ and $y \approx 0.11\;(0.11)$, where $x$ and $y$ are defined as eq.~\eqref{eqn:xydefn}.
	}
\end{table}

Based on this, we performed a numerical analysis to obtain cosmological observables, for a pivot scale $k_{*} = 0.05\,{\rm Mpc}^{-1}$, by considering perturbativity; namely all the coupling constants except the nonminimal couplings should be less than $4\pi$, vacuum stability, and latest Planck result \eqref{eqn:PlanckCO}.
Although a wide range in parameter space generates small tensor-to-scalar ratio, compatible with the SM Higgs inflation, some values of $\alpha$ near the value which source the inflection point give sizeable values of $r$.
Figure~\ref{fig:paramscan} shows the tensor-to-scalar ratio as a function of the mixing angle $\alpha$ for four different values of $m_{s}$.
The grey-dotted lines indicate the parameter region where the spectral index $n_{s}$ becomes larger than the 2$\sigma$ Planck bound, $n_{s}\gtrsim 0.98$.
As expected from the above discussion, for a given value of $m_{s}$, we found that the tensor-to-scalar ratio $r$ becomes small for large values of the mixing angle $\alpha$, while it is possible to obtain large values of $r$ for small values of $\alpha$.
We also found similar behaviors for different sets of model parameters.
As an example, we present two parameter sets which give a sizeable value of $r$ in table~\ref{tab:InfObs}.
For the parameter values used in table~\ref{tab:InfObs}, we found that the conditions \eqref{eqn:condxy} are satisfied.
It is interesting to note that $0.01 \lesssim r \lesssim 0.08$ is difficult to realize in the SFDM model, while $0.08 \lesssim r \lesssim 0.1$ is allowed.
Thus detection of a nonzero tensor-to-scalar ratio in the near future will be an important signal in discriminating different variants of the Higgs inflation.

Let us finally comment on the other constraints on the SFDM model.
Following refs.~\cite{Baek:2011aa,Baek:2012uj}, we consider (i) perturbative unitarity condition, (ii) DM relic density, and (iii) DM direct detection as follows:

\textbf{(i) Perturbative unitarity condition}\\
For $m_{h} \neq m_{s}$, the perturbative unitarity of scattering amplitudes for longitudinal weak gauge bosons gives the following constraint on the mixing angle $\alpha$ for a given dark Higgs mass,
\begin{align}
	\sin^{2}\alpha \leq \frac{1}{m_{s}^{2}-m_{h}^{2}}\left(
		\frac{4\pi\sqrt{2}}{3G_{{\rm F}}} - m_{h}^{2}
	\right)\,.
\end{align}
We find that no severe constraints exist for $m_{s} = [300\,{\rm GeV},600\,{\rm GeV}]$.

\textbf{(ii) DM relic density}\\
In refs.~\cite{Baek:2011aa,Baek:2012uj}, the authors found that the DM relic density, $\Omega_{{\rm CDM}}h^{2} = 0.1198\pm 0.0015$ \cite{Ade:2015xua}, can be satisfied if $m_{\psi} \approx m_{h}/2$ or $m_{\psi} \approx m_{s}/2$, provided that $m_{\psi} < m_{s},m_{h}$.

\begin{figure}[t]
\centering
\includegraphics[width=0.45\textwidth]{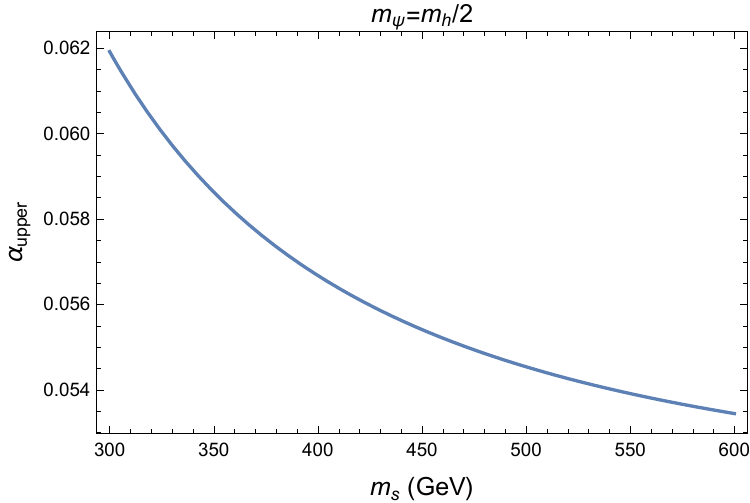}
\quad
\includegraphics[width=0.45\textwidth]{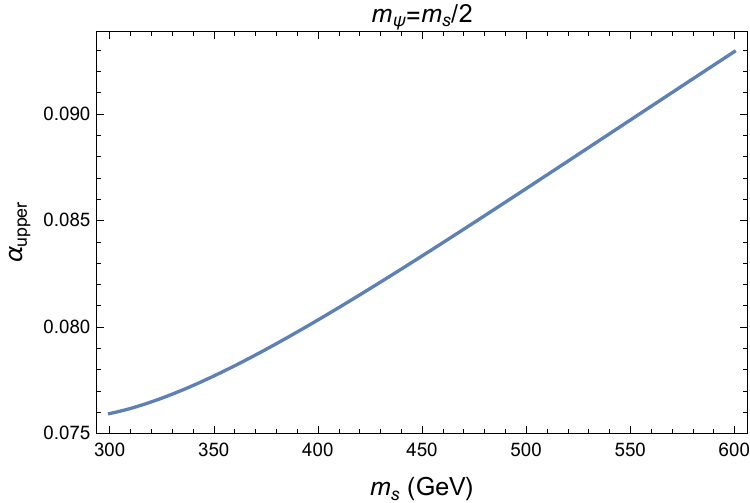}
\caption{\label{fig:LUXbound}
Upper bounds on the mixing angle $\alpha$ from the LUX experiment \cite{Akerib:2016vxi} for the SM Higgs resonance case $m_{\psi}=m_{h}/2$ (left) and the dark Higgs resonance case $m_{\psi}=m_{s}/2$ (right). The SM Higgs resonance case gives more strict bounds on the mixing angle.
Here $\lambda_\psi = 0.3$ is used.
}
\end{figure}

\textbf{(iii) DM direct detection}\\
The upper bound on the spin-independent cross-section of the DM-proton scattering is given from the latest LUX experiment \cite{Akerib:2016vxi}, by
\begin{align}
	\sigma \lesssim 2.2 \times 10^{-46} \, {\rm cm^{2}}
\end{align}
at $m_{\psi} \simeq 50\,{\rm GeV}$.
For $m_{\psi} = \mathcal{O}(10-10^{3})\,{\rm GeV}$, the LUX bound is given by
\begin{align}
	\sigma \lesssim \mathcal{O}(0.1 - 1) 
	\times 10^{-9}
	\, {\rm pb}\,.
\end{align}
If $m_{\psi} \gg m_{p}$, we find \cite{Baek:2011aa,Baek:2012uj}
\begin{align}
	\sigma \simeq 8.56 \times 10^{-9}\,{\rm pb}
	\left(
		\frac{125\,{\rm GeV}}{m_{h}}
	\right)^{4}\left(
		1 - \frac{m_{h}^{2}}{m_{s}^{2}}
	\right)^{2}\left(
		\frac{\lambda_{\psi}\sin\alpha\cos\alpha}{0.1}
	\right)^{2}\,.
\end{align}
For $m_{s}=[300\,{\rm GeV},600\,{\rm GeV}]$, the upper bounds on the mixing angle $\alpha$ are shown in figure~\ref{fig:LUXbound} for $m_{\psi}=m_{h}/2$ and $m_{\psi}=m_{s}/2$ cases.
As shown in figure~\ref{fig:LUXbound}, the case of the SM Higgs resonance, $m_{\psi}=m_{h}/2$ gives more severe constraints on the upper bound on the mixing angle $\alpha$.

In refs. \cite{Baek:2011aa,Baek:2012uj}, the authors found constraints on $\lambda_{\psi}$, $\lambda_{SH}$ and $\lambda_{S}$, considering additionally LEP bound and oblique parameters:
\begin{align}
	0 \leq |\lambda_{\psi}| \lesssim 0.6
	\,,\quad
	-0.2 \lesssim \lambda_{SH} \lesssim 0.4
	\,,\quad
	0 \leq \lambda_{S} \lesssim 0.2\,.
\end{align}
The benchmark point we used throughout this section agrees well with the above constraints.

\begin{figure}[t]
\centering
\includegraphics[width=0.65\textwidth]{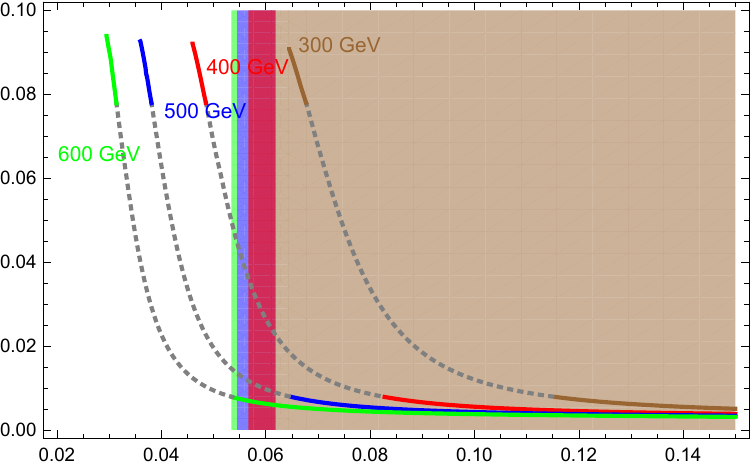}
\caption{\label{fig:rVSalphaFINAL}
Tensor-to-scalar ratio as a function of the mixing angle $\alpha$ for $m_{s}=300$ GeV, 400 GeV, 500 GeV and 600 GeV, with the constraints discussed in the main text. The stringent upper bounds for a given $m_{s}$ comes from the DM physics. The values of the other parameters are the same as in figure~\ref{fig:paramscan}. Color-shaded regions (following the scheme of colored lines) are the excluded regions from the latest LUX experiment, corresponding to different dark Higgs masses.
}
\end{figure}

Figure~\ref{fig:rVSalphaFINAL} shows our final results on the tensor-to-scalar ratio in terms of the mixing angle $\alpha$ for four different dark Higgs masses, considering all the constraints we discussed so far.
Note that small dark Higgs masses, $m_{s}\leq 300$ GeV, are ruled out by the combination of cosmological observations, theoretical constraints, and DM physics.
It is interesting to notice that, for $ 400 \leq m_{s} \leq 600\,{\rm GeV}$, DM physics starts to rule out small values of tensor-to-scalar ratio, favoring $\mathcal{O}(0.1)$ tensor-to-scalar ratio.
We expect to test the Higgs-portal inflation model in the near future as precise measurements of collider/DM/cosmology experiments become available.

\section{Conclusions}
\label{sec:con}
In this paper, we pointed out that a strong connection of masses of the SM Higgs and top quark pole mass in the Higgs inflation is demolished in the presence of Higgs portal interaction, while being consistent with the latest experimental results.
In particular we showed how a large tensor-to-scalar ratio $r \sim 0.08$, which can be probed in the near future experiments, can be achieved in the Higgs inflation without resort to a strong dependence on $M_{t}$.
Using the model of singlet fermion dark matter as a concrete model, we performed a numerical analysis and showed how it is realized.
It is interesting to see that the combination of cosmological observables, theoretical constraints and DM physics rules out small dark Higgs masses and favors $r \sim 0.08$ for $400 \leq m_{s} \leq 600\,{\rm GeV}$, but it is also possible to have a small tensor-to-scalar ratio for very large dark Higgs masses $m_{s} \gtrsim 600\,{\rm GeV}$.
Even though we considered a model in which SM Higgs couples to a real singlet scalar, we expect a similar result for the case where the dark Higgs is charged under a local dark symmetry.
So, given that the Higgs portal interaction is generic in scenarios beyond the SM, playing crucial roles in low energy phenomenology and dark matter physics, where the dark matter is stabilized by a local dark symmetry and thermalized by Higgs portal interaction, we find it amusing that the dark Higgs guarantees the dark matter stability and improves the stability of electroweak vacuum as well as assisting the Higgs inflation at the same time.

\acknowledgments{
JK would like to thank Eibun Senaha, Tommi Tenkanen, and Aditya Aravind for useful discussions.
WIP acknowledges support from the MEC and FEDER (EC) Grants SEV-2014-0398 and FPA2014-54459 and the Generalitat Valenciana under grant PROMETEOII/2013/017. This project has received funding from the European Unions Horizon 2020 research and innovation programme under the Marie Sklodowska-Curie grant Elusives ITN agreement No 674896  and InvisiblesPlus RISE, agreement No 690575.
}

\appendix
\section{Renormalization Group Equations}
\label{apdx:RGEs}
We present the two-loop RG equations of the SFDM model used in our analysis,
following the arguments given in refs.~\cite{DeSimone:2008ei,Lerner:2009xg,Lerner:2011ge} (see also refs.~\cite{Baek:2012uj,Aravind:2015xst}).
The beta functions are given by
\begin{align}\label{eqn:fullRGEs}
  \beta_{\lambda_{H}}
  &=\frac{1}{16\pi^{2}}\left[
    6(1+3s_{h}^{2})\lambda_{H}^{2}
    +12\lambda_{H}\lambda_{t}^{2}
    -6\lambda_{t}^{4}
    -3\lambda_{H}(3g_{2}^{2}+g_{1}^{2})
    +\frac{3}{8}\left(
    2g_{2}^{4}+(g_{2}^{2}+g_{1}^{2})^{2}
    \right)
    +\frac{1}{2}s_{s}^{2}\lambda_{SH}^{2}
    \right]
  \nonumber\\
  &\quad
  +\frac{1}{(16\pi^{2})^{2}}\bigg\{
    \frac{1}{48}\left[
		(912+3s_{h})g_{2}^{6}
		-(290 -s_{h})g_{1}^{2}g_{2}^{4}
		-(560 - s_{h})g_{1}^{4}g_{2}^{2}
		-(380 - s_{h})g_{1}^{6}
	\right]
	\nonumber\\
	&\quad
	+(38 - 8s_{h})\lambda_{t}^{6}
	-\lambda_{t}^{4}\left[
		\frac{8}{3}g_{1}^{2} + 32g_{3}^{2}
		+(12 - 117s_{h} + 108s_{h}^{2})\lambda_{H}
	\right]
	\nonumber\\
	&\quad
	+\lambda_{H}\bigg[
		-\frac{1}{8}(181 + 54s_{h} - 162s_{h}^{2})g_{2}^{4}
		+\frac{1}{4}(3 - 18s_{h} + 54s_{h}^{2})g_{1}^{2}g_{2}^{2}
		+\frac{1}{24}(90 + 377s_{h} + 162s_{h}^{2})g_{1}^{4}
	\nonumber\\
	&\quad
	+(27 + 54s_{h} + 27s_{h}^{2})g_{2}^{2}\lambda_{H}
	+(9 + 18s_{h} + 9s_{h}^{2})g_{1}^{2}\lambda_{H}
	\nonumber\\
	&\quad
	-(48 + 288s_{h} - 324s_{h}^{2} + 624s_{h}^{3} - 324s_{h}^{4})\lambda_{H}^{2}
	\bigg]
	\nonumber\\
	&\quad
	+\lambda_{t}^{2}\left[
		-\frac{9}{4}g_{2}^{4}
		+\frac{21}{2}g_{1}^{2}g_{2}^{2}
		-\frac{19}{4}g_{1}^{4}
		+\lambda_{H}\left(
			\frac{45}{2}g_{2}^{2}
			+\frac{85}{6}g_{1}^{2}
			+80g_{3}^{2}
			-(36 + 108s_{h}^{2})\lambda_{H}
		\right)
	\right]
	\nonumber\\
	&\quad
	-2\lambda_{SH}^{3}
	-5\lambda_{H}\lambda_{SH}^{2}
	-2\lambda_{SH}^{2}\lambda_{\psi}^{2}
    \bigg\}
    \,,\nonumber\\
  \beta_{\lambda_{SH}}
  &=
  \frac{\lambda_{SH}}{16\pi^{2}}
    \left[
    2\left(
    3(1+s_{h}^{2})\lambda_{H}
    +3s_{s}^{2}\lambda_{S}
    +2s_{s}s_{h}\lambda_{SH}
    \right)
    -\left(
    \frac{3}{2}(3g_{2}^{2}+g_{1}^{2})
    -6\lambda_{t}^{2}
    \right)
    +4\lambda_{\psi}^{2}
    \right]
    \nonumber\\
    &\quad
    +\frac{1}{(16\pi^{2})^{2}}\bigg\{
    -\frac{21}{2}\lambda_{SH}^{3}
	- 72\lambda_{SH}^{2}\lambda_{H}
	+\lambda_{SH}^{2}g_{1}^{2}
	-60\lambda_{SH}\lambda_{H}^{2}
	-36\lambda_{SH}^{2}\lambda_{S}
	-30\lambda_{SH}\lambda_{S}^{2}
	-12\lambda_{SH}^{2}\lambda_{t}^{2}
	\nonumber\\
	&\quad
	+3\lambda_{SH}^{2}g_{2}^{2}
	-\frac{145}{16}\lambda_{SH}g_{2}^{4}
	-\frac{27}{2}\lambda_{SH}\lambda_{t}^{4}
	+\frac{557}{48}\lambda_{SH}g_{1}^{4}
	-72\lambda_{SH}\lambda_{H}\lambda_{t}^{2}
	-8\lambda_{SH}^{2}\lambda_{\psi}^{2}
	\nonumber\\
	&\quad
	+24\lambda_{SH}\lambda_{H}g_{1}^{2}
	+72\lambda_{SH}\lambda_{H}g_{2}^{2}
	+40\lambda_{SH}g_{3}^{2}\lambda_{t}^{2}
	+\frac{15}{8}\lambda_{SH}g_{2}^{2}g_{1}^{2}
	\nonumber\\
	&\quad
	+\frac{45}{4}\lambda_{SH}g_{2}^{2}\lambda_{t}^{2}
	+\frac{85}{12}\lambda_{SH}g_{1}^{2}\lambda_{t}^{2}
	-24\lambda_{SH}\lambda_{S}\lambda_{\psi}^{2}
	-2\lambda_{SH}\lambda_{\psi}^{4}
    \bigg\}
    \,,\nonumber\\
  \beta_{\lambda_{S}}
  &=
  \frac{1}{16\pi^{2}}\left[
    \frac{1}{2}(3+s_{h}^{2})\lambda_{SH}^{2}
    +18s_{s}^{2}\lambda_{S}^{2}
    +8\lambda_{\psi}^{2}\lambda_{S}
    -8\lambda_{\psi}^{4}
    \right]
    +\frac{1}{(16\pi^{2})^{2}}\bigg\{
    -8\lambda_{SH}^{3}
	-20\lambda_{SH}^{2}\lambda_{S}
    \nonumber\\
    &\quad
	-204\lambda_{S}^{3}
	-12\lambda_{SH}^{2}\lambda_{t}^{2}
	+12\lambda_{SH}^{2}g_{2}^{2}
	+4\lambda_{SH}^{2}g_{1}^{2}
	-72\lambda_{S}^{2}\lambda_{\psi}^{2}
	+28\lambda_{S}\lambda_{\psi}^{4}
	+64\lambda_{\psi}^{6}
    \bigg\}
    \,,\nonumber\\
    \beta_{\lambda_{\psi}}
  &=\frac{5s_{s}\lambda_{\psi}^{3}}{16\pi^{2}}
    +\frac{\lambda_{\psi}}{(16\pi^{2})^{2}}\left[
    \lambda_{SH}^{2}
	+3\lambda_{S}^{2}
	-\frac{57}{4}\lambda_{\psi}^{4}
	-12\lambda_{S}\lambda_{\psi}^{2}
    \right]
    \,,\nonumber\\
    \beta_{g_{1}}
  &=
  \frac{1}{16\pi^{2}}\left[
  \frac{81+s_{h}}{12}g_{1}^{3}
  \right]
  +\frac{1}{(16\pi^{2})^{2}}\left[
  \frac{199}{18}g_{1}^{5}
	+\frac{9}{2}g_{1}^{3}g_{2}^{2}
	+\frac{44}{3}g_{1}^{3}g_{3}^{2}
	-\frac{17}{6}s_{h}g_{1}^{3}\lambda_{t}^{2}
  \right]
    \,,\nonumber\\
    \beta_{g_{2}}
  &=
  \frac{1}{16\pi^{2}}\left[
    \frac{s_{h}-39}{12}g_{2}^{3}
    \right]
    +\frac{1}{(16\pi^{2})^{2}}\left[
   \frac{3}{2}g_{1}^{2}g_{2}^{3}
	+\frac{35}{6}g_{2}^{5}
	+12g_{2}^{3}g_{3}^{2}
	-\frac{3}{2}s_{h}g_{2}^{3}\lambda_{t}^{2}
    \right]
    \,,\nonumber\\
    \beta_{g_{3}}
  &=
    -\frac{7}{16\pi^{2}}g_{3}^{3}
    +\frac{1}{(16\pi^{2})^{2}}\left[
    \frac{11}{6}g_{1}^{2}g_{3}^{3}
	+\frac{9}{2}g_{2}^{2}g_{3}^{3}
	-26g_{3}^{5}
	-2s_{h}g_{3}^{3}\lambda_{t}^{2}
    \right]
    \,,\nonumber
\end{align}
\begin{align}
	\beta_{\lambda_{t}}
  &=
    \frac{\lambda_{t}}{16\pi^{2}}
    \left[
    \left(
    \frac{23}{6}+\frac{2}{3}s_{h}
    \right)\lambda_{t}^{2}
    -\left(
    8g_{3}^{2}
    +\frac{9}{4}g_{2}^{2}
    +\frac{17}{12}g_{1}^{2}
    \right)
    \right]
    \nonumber\\
    &\quad
    +\frac{\lambda_{t}}{(16\pi^{2})^{2}}
    \bigg[
		-\frac{23}{4}g_{2}^{4}
		-\frac{3}{4}g_{1}^{2}g_{2}^{2}
		+\frac{1187}{216}g_{1}^{4}
		+9g_{2}^{2}g_{3}^{2}
		+\frac{19}{9}g_{1}^{2}g_{3}^{2}
		-108g_{3}^{4}
		\nonumber\\
		&\quad
		+\left(
			\frac{225}{16}g_{2}^{2}
			+\frac{131}{16}g_{1}^{2}
			+36g_{3}^{2}
		\right)s_{h}\lambda_{t}^{2}
		+6\left(
			-2s_{h}^{2}\lambda_{t}^{4}
			-2s_{h}^{3}\lambda_{t}^{2}\lambda_{H}
			+s_{h}^{2}\lambda_{h}^{2}
		\right)
		+\frac{1}{4}\lambda_{SH}^{2}
	\bigg]
    \,,
\end{align}
where $\lambda_{t}$ is the top Yukawa coupling, $g_{1,2,3}$ are the SM gauge couplings, and
\begin{align}
  s_{\phi} = \frac{1+\xi_{\phi}\phi^{2}/M_{{\rm P}}^{2}}{1+(1+6\xi_{\phi})\xi_{\phi}\phi^{2}/M_{{\rm P}}^{2}}
  \,,\qquad
  \phi= \{h,s\}
\end{align}
is the suppression factor.
Note that, at two-loop level, we neglected the suppression factor for the new physics contribution while taking it into account for the SM parameters.
For the nonminimal couplings we use one-loop beta functions,
\begin{align}
  16\pi^{2}\beta_{\xi_{h}}
  &=
    \left[
    6(1+s_{h})\lambda_{H}
    +6y_{t}^{2}
    -\frac{3}{2}(3g_{2}^{2}+g_{1}^{2})
    \right]\left(
    \xi_{h}+\frac{1}{6}
    \right)
    +s_{s}\lambda_{SH}\left(
    \xi_{s}+\frac{1}{6}
    \right)
    \,,\nonumber\\
  16\pi^{2}\beta_{\xi_{s}}
  &=
    (3+s_{h})\lambda_{SH}\left(
    \xi_{h}+\frac{1}{6}
    \right)
    +6s_{s}\lambda_{S}\left(
    \xi_{s}+\frac{1}{6}
    \right)
    \,.
\end{align}
Note that the beta functions are defined by, for any coupling constant $g$ except the nonminimal couplings,
\begin{align}
  \beta_{g} = (1+\gamma)\frac{dg}{dt}
\end{align}
for inflation occuring along the SM Higgs field direction, and
\begin{align}
  \beta_{g} = \frac{dg}{dt}
\end{align}
for inflation along $s$-field direction, where
\begin{align}
  \gamma &= -\frac{1}{16\pi^{2}}\left(
  \frac{9}{4}g_{2}^{2}+\frac{3}{4}g_{1}^{2}-3\lambda_{t}^{2}
  \right)
  +\frac{1}{(16\pi^{2})^{2}}
  \bigg[
		\frac{271}{32}g_{2}^{4}
		-\frac{9}{16}g_{1}^{2}g_{2}^{2}
		-\frac{431}{96}s_{h}g_{1}^{4}
  \nonumber\\
  &\quad
  -\frac{5}{2}\left(
	\frac{9}{4}g_{2}^{2}
	+\frac{17}{12}g_{1}^{2}
	+8g_{3}^{2}
	\right)\lambda_{t}^{2}
	+\frac{27}{4}s_{h}\lambda_{t}^{4}
	-6s_{h}^{3}\lambda_{H}^{2}
	+\frac{1}{4}\lambda_{SH}^{2}
	\bigg]
	\,.
\end{align}
is the Higgs field anomalous dimension and $t = \ln(\mu/M_{t})$.



\begin{thebibliography}{99}
\bibitem{Bezrukov:2007ep} 
  F.~L.~Bezrukov and M.~Shaposhnikov,
  Phys.\ Lett.\ B {\bf 659}, 703 (2008)
  [arXiv:0710.3755 [hep-th]].

\bibitem{Germani:2010gm} 
  C.~Germani and A.~Kehagias,
  Phys.\ Rev.\ Lett.\  {\bf 105}, 011302 (2010)
  [arXiv:1003.2635 [hep-ph]].

\bibitem{Ade:2015xua} 
  P.~A.~R.~Ade {\it et al.} [Planck Collaboration],
  arXiv:1502.01589 [astro-ph.CO].

\bibitem{Ade:2015lrj} 
  P.~A.~R.~Ade {\it et al.} [Planck Collaboration],
  arXiv:1502.02114 [astro-ph.CO].

\bibitem{DeSimone:2008ei} 
  A.~De Simone, M.~P.~Hertzberg and F.~Wilczek,
  Phys.\ Lett.\ B {\bf 678}, 1 (2009)
  doi:10.1016/j.physletb.2009.05.054
  [arXiv:0812.4946 [hep-ph]].
    
\bibitem{Hamada:2014iga} 
  Y.~Hamada, H.~Kawai, K.~-y.~Oda and S.~C.~Park,
  arXiv:1403.5043 [hep-ph].
  
\bibitem{Bezrukov:2014bra} 
  F.~Bezrukov and M.~Shaposhnikov,
  arXiv:1403.6078 [hep-ph].

\bibitem{Germani:2014hqa} 
  C.~Germani, Y.~Watanabe and N.~Wintergerst,
  JCAP {\bf 1412}, no. 12, 009 (2014)
  doi:10.1088/1475-7516/2014/12/009
  [arXiv:1403.5766 [hep-ph]].

\bibitem{Ade:2014xna} 
  P.~A.~R.~Ade {\it et al.}  [BICEP2 Collaboration],
  arXiv:1403.3985 [astro-ph.CO].

\bibitem{Bezrukov:2014ipa} 
  F.~Bezrukov, J.~Rubio and M.~Shaposhnikov,
  Phys.\ Rev.\ D {\bf 92}, no. 8, 083512 (2015)
  doi:10.1103/PhysRevD.92.083512
  [arXiv:1412.3811 [hep-ph]].

\bibitem{Buttazzo:2013uya}
  D.~Buttazzo, G.~Degrassi, P.~P.~Giardino, G.~F.~Giudice, F.~Sala, A.~Salvio and A.~Strumia,
  JHEP {\bf 1312} (2013) 089
  doi:10.1007/JHEP12(2013)089
  [arXiv:1307.3536 [hep-ph]].

\bibitem{Degrassi:2012ry} 
  G.~Degrassi, S.~Di Vita, J.~Elias-Miro, J.~R.~Espinosa, G.~F.~Giudice, G.~Isidori and A.~Strumia,
  JHEP {\bf 1208}, 098 (2012)
  doi:10.1007/JHEP08(2012)098
  [arXiv:1205.6497 [hep-ph]].

\bibitem{Allison:2013uaa} 
  K.~Allison,
  JHEP {\bf 1402}, 040 (2014)
  doi:10.1007/JHEP02(2014)040
  [arXiv:1306.6931 [hep-ph]].

\bibitem{Haba:2014zda} 
  N.~Haba and R.~Takahashi,
  arXiv:1404.4737 [hep-ph].

\bibitem{Ballesteros:2015iua} 
  G.~Ballesteros and C.~Tamarit,
  JHEP {\bf 1509}, 210 (2015)
  doi:10.1007/JHEP09(2015)210
  [arXiv:1505.07476 [hep-ph]].

\bibitem{Lerner:2009xg} 
  R.~N.~Lerner and J.~McDonald,
  Phys.\ Rev.\ D {\bf 80}, 123507 (2009)
  doi:10.1103/PhysRevD.80.123507
  [arXiv:0909.0520 [hep-ph]].
  
\bibitem{Lerner:2011ge} 
  R.~N.~Lerner and J.~McDonald,
  Phys.\ Rev.\ D {\bf 83}, 123522 (2011)
  doi:10.1103/PhysRevD.83.123522
  [arXiv:1104.2468 [hep-ph]].
  
\bibitem{Ballesteros:2016xej} 
  G.~Ballesteros, J.~Redondo, A.~Ringwald and C.~Tamarit,
  arXiv:1610.01639 [hep-ph].
  
\bibitem{Tenkanen:2016idg} 
  T.~Tenkanen, K.~Tuominen and V.~Vaskonen,
  arXiv:1606.06063 [hep-ph].
  
\bibitem{Aravind:2015xst} 
  A.~Aravind, M.~Xiao and J.~H.~Yu,
  Phys.\ Rev.\ D {\bf 93}, no. 12, 123513 (2016)
  doi:10.1103/PhysRevD.93.123513
  [arXiv:1512.09126 [hep-ph]].

\bibitem{Kawai:2014gqa} 
  S.~Kawai and J.~Kim,
  Phys.\ Rev.\ D {\bf 91}, no. 4, 045021 (2015)
  doi:10.1103/PhysRevD.91.045021
  [arXiv:1411.5188 [hep-ph]].

\bibitem{Kawai:2015ryj} 
  S.~Kawai and J.~Kim,
  Phys.\ Rev.\ D {\bf 93}, no. 6, 065023 (2016)
  doi:10.1103/PhysRevD.93.065023
  [arXiv:1512.05861 [hep-ph]].
  

\bibitem{hur_ko}
T.~Hur and P.~Ko,
  Phys.\ Rev.\ Lett.\  {\bf 106}, 141802 (2011)
  [arXiv:1103.2571 [hep-ph]].

\bibitem{Baek:2011aa} 
  S.~Baek, P.~Ko and W.~-I.~Park,
  JHEP {\bf 1202}, 047 (2012)
  [arXiv:1112.1847 [hep-ph]].

\bibitem{Baek:2012uj} 
  S.~Baek, P.~Ko, W.~-I.~Park and E.~Senaha,
  JHEP {\bf 1211}, 116 (2012)
  [arXiv:1209.4163 [hep-ph]].

\bibitem{Baek:2012se} 
  S.~Baek, P.~Ko, W.~-I.~Park and E.~Senaha,
  JHEP {\bf 1305}, 036 (2013)
  [arXiv:1212.2131 [hep-ph]].

\bibitem{Baek:2013qwa} 
  S.~Baek, P.~Ko and W.~-I.~Park,
  JHEP {\bf 1307}, 013 (2013)
  [arXiv:1303.4280 [hep-ph]].

\bibitem{Baek:2013dwa} 
  S.~Baek, P.~Ko and W.~-I.~Park,
  arXiv:1311.1035 [hep-ph].

\bibitem{Baek:2014goa} 
  S.~Baek, P.~Ko, W.~-I.~Park and Y.~Tang,
  arXiv:1402.2115 [hep-ph].

\bibitem{Ko:2014nha} 
  P.~Ko and Y.~Tang,
  arXiv:1402.6449 [hep-ph].

\bibitem{Ko:2014bka} 
  P.~Ko and Y.~Tang,
  arXiv:1404.0236 [hep-ph].



\bibitem{Kinney:2016qyl} 
  W.~H.~Kinney,
  arXiv:1606.00672 [astro-ph.CO].

\bibitem{Olive:2016xmw} 
  C.~Patrignani {\it et al.} [Particle Data Group Collaboration],
  Chin.\ Phys.\ C {\bf 40}, no. 10, 100001 (2016).
  doi:10.1088/1674-1137/40/10/100001

\bibitem{George:2015nza} 
  D.~P.~George, S.~Mooij and M.~Postma,
  JCAP {\bf 1604}, no. 04, 006 (2016)
  doi:10.1088/1475-7516/2016/04/006
  [arXiv:1508.04660 [hep-th]].

\bibitem{Sher:1988mj}
  M.~Sher,
  Phys.\ Rept.\  {\bf 179} (1989) 273.
  doi:10.1016/0370-1573(89)90061-6

\bibitem{Stewart:1993bc} 
  E.~D.~Stewart and D.~H.~Lyth,
  Phys.\ Lett.\ B {\bf 302}, 171 (1993)
  doi:10.1016/0370-2693(93)90379-V
  [gr-qc/9302019].

\bibitem{Liddle:1994dx} 
  A.~R.~Liddle, P.~Parsons and J.~D.~Barrow,
  Phys.\ Rev.\ D {\bf 50}, 7222 (1994)
  doi:10.1103/PhysRevD.50.7222
  [astro-ph/9408015].

\bibitem{Leach:2002ar} 
  S.~M.~Leach, A.~R.~Liddle, J.~Martin and D.~J.~Schwarz,
  Phys.\ Rev.\ D {\bf 66}, 023515 (2002)
  doi:10.1103/PhysRevD.66.023515
  [astro-ph/0202094].

\bibitem{Gorbunov:2012ns} 
  D.~Gorbunov and A.~Tokareva,
  JCAP {\bf 1312}, 021 (2013)
  doi:10.1088/1475-7516/2013/12/021
  [arXiv:1212.4466 [astro-ph.CO]].

\bibitem{Kniehl:2016enc} 
  B.~A.~Kniehl, A.~F.~Pikelner and O.~L.~Veretin,
  Comput.\ Phys.\ Commun.\  {\bf 206}, 84 (2016)
  doi:10.1016/j.cpc.2016.04.017
  [arXiv:1601.08143 [hep-ph]].

\bibitem{Kniehl:2015nwa} 
  B.~A.~Kniehl, A.~F.~Pikelner and O.~L.~Veretin,
  Nucl.\ Phys.\ B {\bf 896}, 19 (2015)
  doi:10.1016/j.nuclphysb.2015.04.010
  [arXiv:1503.02138 [hep-ph]].

\bibitem{Liddle:1993fq} 
  A.~R.~Liddle and D.~H.~Lyth,
  Phys.\ Rept.\  {\bf 231}, 1 (1993)
  doi:10.1016/0370-1573(93)90114-S
  [astro-ph/9303019].

  
\bibitem{Akerib:2016vxi} 
  D.~S.~Akerib {\it et al.},
  arXiv:1608.07648 [astro-ph.CO].
  
\end{thebibliography}
\end{document}